\documentstyle[12pt,epsf]{article}
\newcommand{\Z}{{Z \!\!\! Z}}
\newcommand{\dual}{\mbox{}^{\ast}}
\newcommand{\LL}{{I\!\! L}}

\newcommand{\eq}[1]{(\ref{#1})}

\newcommand{\beq}{\begin{equation}}
\newcommand{\eeq}{\end{equation}}
\newcommand{\beqn}{\begin{eqnarray}}
\newcommand{\eeqn}{\end{eqnarray}}

\newcommand{\cD}{{\cal D}}

\def\cC{{\cal C}}

\newcommand{\cZ}{{\cal Z}}

\newcommand{\intpiD}{\int\limits_{-\pi}^{+\pi} {\cD}}

\newcommand{\intinfD}{\int\limits_{-\infty}^{+\infty} {\cD}}
\newcommand{\expb}[1]{\exp\left\{ #1 \right\} }
\newcommand{\CK}[1]{\mbox{\scriptsize c}_{\mbox{$\scriptstyle #1$}}}
\newcommand{\nsum}[2]{\sum_{ #1(\CK{#2}) \in \Z }}
\newcommand{\ndsum}[2]{\sum_{\stackrel{\scriptstyle #1(\CK{#2}) \in \Z}
{\delta #1=0}}}

\newcommand{\nddsum}[2]{\sum_{\stackrel{\scriptstyle \dual #1(\dual\CK{#2})
\in \Z} {\delta \dual #1=0}}}
\newcommand{\dd}{\mbox{d}}

\def\dd{{\rm d}}

\def\NP{ Nucl.~Phys.}
\def\PL{ Phys.~Lett.}
\def\PR{ Phys.~Rev.}
\def\PRL{ Phys.~Rev.~Lett.}

\begin{document}
\bibliographystyle{bibstand}
\sloppy
\date{}

\title{
\vspace{-1.5cm}
\begin{flushright}
\begin{large}
ITEP-TH-29/96\\
\end{large}
\end{flushright}
\vspace{1.5cm}
Analytical and Numerical Study of the
Aharonov--Bohm Effect in $3D$ and $4D$ Abelian Higgs
Model\thanks{Talk given at the 9th International Seminar
``Quarks-96'', May 1996, Yaroslavl, Russia.}}

\author{M.N.~Chernodub, F.V.~Gubarev and M.I.~Polikarpov \\{\sl
ITEP, B.Cheremushkinskaya 25, Moscow, 117259, Russia}}

\maketitle

\begin{abstract}
We discuss the Aharonov--Bohm effect in three and four dimensional
non--compact lattice Abelian Higgs model. We show analytically that
this effect leads to the long--range Coulomb interaction of the
charged particles, which is confining in three dimensions. The
Aharonov--Bohm effect is found in numerical calculations
in $3D$ Abelian Higgs model.
\end{abstract}

\section{Introduction}

It is well known that the four dimensional Abelian Higgs model has
classical solutions called Abrikosov--Nielsen--Olesen
strings~\cite{AbNiOl}. These strings carry quantized magnetic flux
and the wave function of the charged particle which is scattered on
this string acquires additional phase. The shift in the phase is the
physical effect which is the field--theoretical analog of the
quantum--mechanical Aharonov--Bohm effect \cite{AhBo59}:  strings
play the role of solenoids, which scatter charged particles.
Topological long--range Aharonov--Bohm interaction
between the strings and particles was discussed in papers
\cite{ABE,threeA,PoWiZu93}.

The three dimensional Abelian Higgs model has the particle--like
solutions usually called Abrikosov vortices.  Analogously to the four
dimensional case, the charged particle which is scattered on the
vortex acquire the additional Aharonov--Bohm phase.

In Section~2 we study the interaction of the charged particles in the
non--compact lattice Abelian Higgs model.  We show that the
Aharonov--Bohm effect gives rise to the long--range Coulomb--like
interaction between test particles. In the three dimensional case the
induced potential is confining, since the Coulomb interaction rises
logarithmically.

In Section~3 we present the results of the numerical calculations in
the $3D$ lattice Abelian Higgs model with the non--compact gauge
field. Our numerical results show the existence of the Aharonov--Bohm
effect in this model.

\section{Potential Induced by Aharonov--Bohm Effect}

The partition function of the $D$ dimensional non--compact Abelian
Higgs model on the lattice is (we are using the formalism of the
differential forms on the lattice, see \cite{PoWiZu93,Intro} for the
brief introduction):

\beq
  \cZ = \intinfD A \intpiD \varphi \nsum{l}{1} \expb{
  - \beta \|\dd A\|^2
  - \gamma \| \dd\varphi + 2\pi l - N A \|^2}\,, \label{NC}\\
\eeq
where $A$ is the non--compact gauge field, $\varphi$ is the phase of
the Higgs field and $l$ is the integer--valued one--form. For
simplicity we consider the limit of the infinite Higgs boson mass,
then the radial part of the Higgs field is frozen and we use the
Villain form of the action.

One can rewrite \cite{PoWiZu93} the integral \eq{NC} as the sum over
the closed vortex ($D=3$) or string ($D=4$) trajectories using the
analogue of Berezinski--Kosterlitz--Thauless (BKT) transformation
\cite{BKT}:

\beq
  \cZ \propto \cZ^{BKT} = const. \, \nddsum{\sigma}{D-2}
  \expb{ - 4 \pi^2 \gamma
  \left(\dual \sigma, {(\Delta + m^2)}^{-1} \dual \sigma \right)}
  \,, \label{TD}
\eeq
where $m^2 = N^2 \gamma \slash \beta$ is the $classical$ mass of the
vector boson $A$. Closed currents $\dual \sigma$ which are defined on
the dual lattice represent the vortex trajectory ($D=3$) or the
string world sheet ($D=4$). It can be easily seen from eq.\eq{TD}
that the currents $\dual \sigma$ interact with each other through the
Yukawa forces.

Using the same transformation for the quantum average of the Wilson
loop $W_M(\cC) = \exp\{ i M (A,j_\cC) \}$ leads to the following
formula~\cite{threeA,PoWiZu93,Trento}:

\beqn
 {<W_M(\cC)>}_N = \frac{1}{\cZ^{BKT}} \nddsum{\sigma}{D-2}
 \exp\biggl\{- 4\pi^2\gamma \left(\dual \sigma,{\left(\Delta +
 m^2\right)}^{-1}\dual \sigma \right) \nonumber\\
 - \frac{M^2}{4 \gamma}( j_\cC,(\Delta + m^2)^{-1} j_\cC) - 2 \pi
 i \frac{M}{N} \left(\dual j_\cC,{\left(\Delta + m^2\right)}^{-1} \dd
 \dual \sigma \right) + 2 \pi i \frac{M}{N} \LL\left(\dual \sigma,
 j_\cC\right)\biggr\}\,.
 \label{wl}
\eeqn
The first three terms in this expression are short--range Yukawa
forces between defects and particles. The last long--range term has
the topological origin: $\LL (\dual \sigma,j_\cC)$ is the linking
number between the world trajectories of the defects $\dual \sigma$
and the Wilson loop~$j_\cC$:

\beq
       \LL(\dual \sigma,j_\cC) = (\dual j_\cC, {\Delta}^{-1} \dd \dual
       \sigma)\,.  \label{Ll}
\eeq
In three (four) dimensions the trajectory of the vortex (string)
$\dual \sigma$ is a closed loop (surface) and the linking number
$\LL$ is equal to the number of points at which the loop $j_\cC$
intersects the two (three) dimensional volume bounded by the loop
(surface)~$\dual \sigma$. The equation \eq{Ll} is the lattice
analogue of the Gauss formula for the linking number. This
topological interaction corresponds to the Aharonov--Bohm effect in
the field theory~\cite{ABE,threeA,PoWiZu93,Trento}.

In the limit

\beqn
N^2 \gamma \gg 1 \gg \beta\,,
\label{limit}
\eeqn
the partition function is:

\beqn
  \cZ^{BKT}_0 = \nddsum{\sigma}{D-2}
  \exp\biggl\{- \frac{4 \pi^2 \beta}{N^2}
  {||\dual \sigma||}^2 \biggr\}\,.
 \label{TD2}
\eeqn
In the corresponding three (four) dimensional continuum theory the
term ${||\dual \sigma||}^2$ is proportional to the length
(area) of the trajectory $\dual \sigma$, therefore in the limit
\eq{limit} the vortices (strings) are free. The quantum average
\eq{wl} of the Wilson loop in the limit \eq{limit} is:

\beqn
 {<W_M(\cC)>}_N = \frac{1}{\cZ^{BKT}_0} \nddsum{\sigma}{D-2}
 \exp\biggl\{- \frac{4 \pi^2 \beta}{N^2} {||\dual \sigma||}^2 +
 2 \pi i \frac{M}{N} \LL\left(\dual \sigma,j_\cC\right) \biggr\}\,.
 \label{wl2}
\eeqn
This formula describes the Aharonov--Bohm interaction of the free
vortices (strings) carrying the flux $\frac{2 \pi}{N}$ with the test
particle of the charge $M$. Therefore the interaction between the
charged particles is due to Aharonov--Bohm effect $only$.

In order to get the explicit expression for ${<W_M(\cC)>}_N$ it is
convenient to rewrite eq.\eq{wl2} in the dual
representation~\cite{Trento}:

\beqn
  \cZ^{BKT}_0 \propto \cZ^{dual}_0 =
  \ndsum{l}{1} \, \exp \biggl\{ - \frac{N^2}{4 \beta}
  {||l||}^2 \biggr\}\,.
  \label{dualZ}
\eeqn

The quantum average \eq{wl2} in the dual representation is:

\beqn
  {<W_M(\cC)>}_N = \frac{1}{\cZ^{dual}_0}
  \ndsum{l}{1} \, \exp \left\{ - \frac{N^2}{4 \beta}
  \left(l - \frac{M}{N} j_\cC, \Delta^{-1}
  \left(l - \frac{M}{N} j_\cC \right) \right) \right\}\,.
  \label{dualWL}
\eeqn
Using the saddle point approximation ($N^2 \slash \beta \gg 1$), we
get in the leading order:

\beq
  {<W_M(\cC)>}_N = const. \, \exp \left\{ -
  \kappa^{(0)}(M, N; \beta)\,
  \left(j_\cC, \Delta^{-1} j_\cC\right) \right\}\,,
  \label{wl3}
\eeq
where

\beq
    \kappa^{(0)}(M, N; \beta) = \frac{q^2 N^2}{4 \beta}\,,
    \label{kappa1}
\eeq
and

\beqn
  q = \min_{K \in \Z} |\frac{M}{N} - K|\,,
  \label{q}
\eeqn
$q$ is the distance between the ratio $M \slash N$ and a nearest
integer number. Expressions (\ref{wl3}--\ref{q}) depend on the
fractional part of $M \slash N$, this is the consequence of the
Aharonov--Bohm effect. Interaction of the testing charges is absent
if $q = 0$ ($M \slash N$ is integer), this corresponds to the
complete screening of the charge $M$ by the charge $N$ of the Higgs
bosons.

Let us consider the product of two Polyakov lines: $W_M(\cC) =
L^+_M(0) \cdot L_M(R)$. Then $(j_\cC, \Delta^{-1} j_\cC) = 2 T \,
\Delta^{-1}_{(D-1)}(R)$, where $\Delta^{-1}_{(D)}(R)$ is the
$D$--dimensional massless lattice propagator, and \eq{wl3} is reduced
to:

\beqn
  {<L^+_M(0) L_M(R)>}_N =
  const. \, \exp \left\{ - 2 \kappa^{(0)}(M, N; \beta) \,
  T \, \Delta^{-1}_{(2)}(R)\right\}.
  \label{wlR}
\eeqn
At large $R$ the propagator $\Delta^{-1}_{(2)}(R)$ rises
logarithmically: $\Delta^{-1}_{(2)}(R) = \frac{c_3}{2} \ln R +
\dots$, and the propagator $\Delta^{-1}_{(3)}(R)$ is inversely
proportional to $R$: $\Delta^{-1}_{(3)}(R) = \frac{c_4}{2} R^{-1} +
\dots$, where $c_3$ and $c_4$ are some numerical constants. Thus, the
Aharonov--Bohm effect in three and four dimensional Abelian Higgs
model gives rise to the following long--range potentials:

\beqn
  V_{D=3}(R) = c_3 \, \kappa^{(0)}(M, N; \beta) \cdot \ln R
  + O\left( R^{-1} \right)\,,\label{V}\\
  V_{D=4}(R) = c_4 \, \kappa^{(0)}(M, N; \beta) \cdot \frac{1}{R}
  + O\left( R^{-2} \right)\,.
\eeqn

A $3D$ vortex model in the continuum was discussed in paper
\cite{Sa79} and it was shown that the potential has the form
$V^{c}(R) \approx const.\, q^2 \psi_0 \, \ln \frac{R}{R_0}$, where
$\psi_0$ is proportional to the vortex condensate, and $R_0$ is of
the order of the vortex width. This expression and our lattice result
\eq{V} are in the agreement.

Note, that in paper \cite{CaDaGr} the confinement of the fractionally
charged particles in the two dimensional Abelian Higgs model was
found. Using the methods which are described in this Section it can
be easily shown that the confinement in this model is due to the
Aharonov--Bohm effect in two dimensions.

\section{Numerical Calculations}

Now we present the results of the numerical calculation of the
potential between the test particles with the charge $M$ in the
three dimensional Abelian Higgs model, the charge of the Higgs boson
is $N=6$. The action of the model is chosen in the Wilson form:
$S[A,\varphi] = \beta {||\dd A||}^2 - \gamma \sum_l \cos {(\dd
\varphi + N A)}_l$. In our calculations we use the standard
Monte--Carlo method.  The simulations are performed on the lattice of
the size $L^2 \times L_t$, $L=16$, $L_t = 2,4$ for the charges $M =
1, \dots, N$.

We fit the numerical data for ${<L^+_M(0) L_M(R)>}_N$ by the formula:

\beq
  \ln \, {<L^+_M(0) L_M(R)>}_N = 2 \, \kappa^{num} \cdot
  T \cdot \Delta^{-1}_{(2)}(R) + C^{num}\,,
  \label{fit}
\eeq
where $\kappa^{num}$ and $C^{num}$ are numerical fitting parameters.
It turns out that the numerical data for $\kappa^{num}(M)$ are well
described by the formula \eq{fit}.

\begin{figure}[bth]
\vspace{-3.5cm}
\centerline{
\epsfxsize=0.65\textwidth
\epsfbox{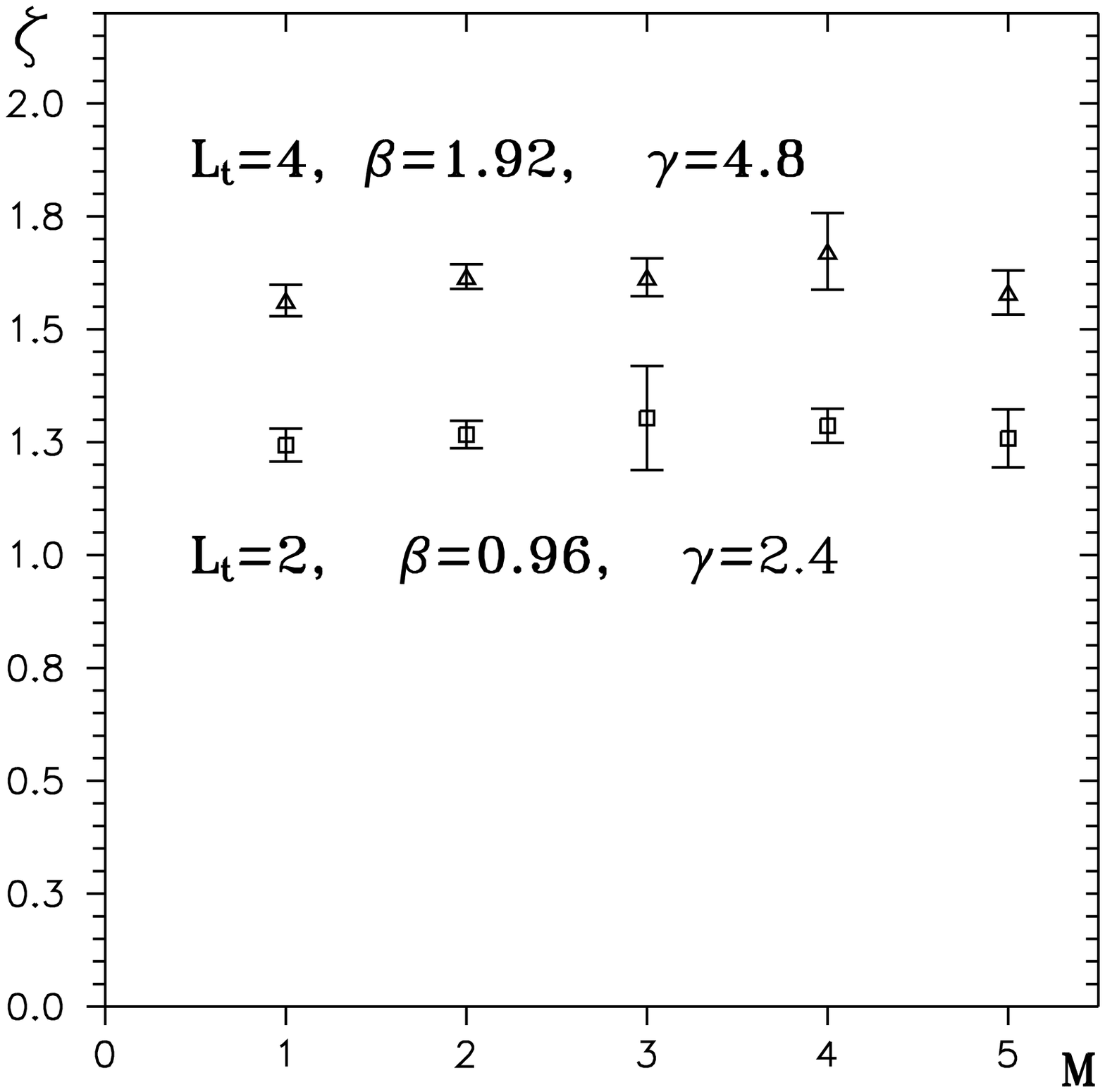}
}
\vspace{1.2cm}
\centerline{Fig.~1: The ratios $\zeta(M)$.}
~
\vspace{.2cm}
\end{figure}

We present on the Fig.~1 the ratios $\zeta(M) = \kappa^{num}(M) /
\kappa^{(0)}(M)$ for $\beta = 0.96$, $\gamma = 2.4$, $L_t = 2$;
$\beta = 1.92$, $\gamma = 4.8$, $L_t = 4$ and $M=1, \dots, 5$, where
$\kappa^{(0)}$ is the semiclassical contribution to $\kappa$,
eq.\eq{kappa1}. It is seen that $\zeta$ for different parameters of
the model does not depend on the charge $M$ (if $M/N$ is not
integer). This result is very interesting since it shows that the
coefficient $\kappa^{num}$ is proportional to $q^2$.  This fact means
that the interaction of the tested particles is due to the
Aharonov--Bohm effect even in the region of the parameters where the
asymptotic expressions (\ref{wl3}--\ref{q}) are not
valid\footnote{Since the usual Coulomb interaction of the test
particles with the charge $M$ is proportional to $M^2$ (and not to
$q^2$), the proportionality of $\kappa^{num}$ to $q^2$ also means
that the Coulomb interaction of the tested particles is small at the
considered values of the parameters of the model.}. The difference
of $\zeta$ from unity can be explained as the renormalization of
$\kappa^{(0)}$ by quantum corrections.

The coefficient $\kappa^{num}$ for $M=N=6$ is zero within the
numerical errors. This result is in agreement with Aharonov--Bohm
nature of the induced potential: the test charge $M$ is screened by
the charge $N$ of the Higgs bosons, therefore the Aharonov--Bohm
effect is absent and the induced potential $V(R)$ has no long--range
terms.

\section*{Conclusion and Acknowledgments}

In this talk we show analytically and numerically that the
Aharonov--Bohm interaction between the topological defects and the
charged particles induce the long--range interaction between the
particles. The induced interaction is of Coulomb type in the
cases when the topological defects are free vortices ($D=3$) and
non--interacting strings ($D=4$). Due to the long--ranged nature of
the induced potential the Aharonov--Bohm effect may play a role in
the dynamics of colour confinement in nonabelian gauge
theories~\cite{ChPoZu94}.

This work is supported by the JSPS Program on Japan -- FSU scientists
collaboration, by the Grants INTAS-94-0840, INTAS-94-2851, and by
Grant No. 96-02-17230a, financed by the Russian Foundation for
Fundamental Sciences.

\end{document}